\documentclass[10pt,twocolumn]{article}

\usepackage{graphicx}
\usepackage{amsfonts}
\usepackage{amsmath}
\usepackage{amssymb}
\usepackage{mathrsfs} 
\usepackage{lipsum}
\usepackage{color}
\usepackage{siunitx}
\usepackage{authblk}
\usepackage[margin=2cm]{geometry}
\usepackage{mathtools, cuted}

\usepackage{bm}
\usepackage{amsmath}

\newcommand{\Bo}{\mathrm{Bo}}
\newcommand{\Tsf}{T_{\mathrm{sf}}}
\newcommand{\Tr}{T_{\mathrm{r}}}
\newcommand{\Qair}{Q_{\mathrm{air}}}
\newcommand{\cmc}{\mathrm{cmc}}
\newcommand{\uth}{u_{\theta}}
\newcommand{\uthbar}{\bar{u}_{\theta}}
\newcommand{\dd}{\mathrm{d}}
\newcommand{\thmin}{\theta_{\mathrm{min}}}
\newcommand{\thrupt}{\theta_{\mathrm{rupt}}}

\newcommand{\milli}{\mathrm{m}}
\newcommand{\centi}{\mathrm{c}}
\newcommand{\meter}{\mathrm{m}}
\newcommand{\per}{\mathrm{/}}
\newcommand{\second}{\mathrm{s}}

\newcommand{\micro}{\mu}

\newcommand{\mole}{\mathrm{mol}}

\newcommand{\liter}{\mathrm{L}}
\newcommand{\mega}{\mathrm{M}}
\newcommand{\nano}{\mathrm{n}}
\newcommand{\ohm}{\Omega}
\newcommand{\newton}{\mathrm{N}}
\newcommand{\minute}{\mathrm{min}}

\title{Life and death of not so ``bare'' bubbles}

\author[1]{Lor\`ene Champougny}
\author[1]{Matthieu Roch\'{e}} 
\author[1]{Wiebke Drenckhan}
\author[1]{Emmanuelle. Rio}

\affil[1] {Universit\'e Paris-Sud, Laboratoire de Physique des Solides, UMR8502, Orsay, F-91405}

\date{\today}

\begin{document}

\twocolumn[
    \begin{@twocolumnfalse}
        \maketitle
\begin{abstract}
In this paper, we investigate how the drainage and rupture of surfactant-stabilised bubbles floating at the surface of a liquid pool depend on the concentration of surface-active molecules in water. Drainage measurements at the apex of bubbles indicate that the flow profile is increasingly plug-like as the surfactant concentration is decreased from several times the critical micellar concentration (cmc) to just below the cmc. High-speed observations of bubble bursting reveal that the position at which a hole nucleates in the bubble cap also depends on the surfactant concentration. On average, the rupture is initiated close to the bubble foot for low concentrations (< cmc) while its locus moves towards the top of the bubble cap as the concentration increases above the cmc. In order to explain this transition, we propose that marginal regeneration may be responsible for bubble rupture at low concentrations but that bursting at the apex for higher concentrations is driven by gravitational drainage.           
\end{abstract}
   \end{@twocolumnfalse}]

\section{Introduction}
%
Numerous natural and industrial processes involve the presence of bubbles which rise to a liquid surface, where they drain and finally rupture. For example, despite their small size compared to typical geophysical scales, bubbles floating at the surface of the oceans can have surprisingly large effects at the scale of the entire planet by mediating mass transfert between the sea and the atmosphere. The tiny droplets expelled upon bubble bursting were observed to contribute significantly to the global production of seaspray aerosols, which in turn affects the climate on Earth\cite{Monahan2001}. Conversely, bubble bursting can also inject surface material into the water column\cite{Feng2014}. At a smaller scale, the aerosols produced upon bubble rupture at the surface of gasified beverages were shown to have a strong impact on the drinker's sensations since they contain most of the flavors\cite{Liger-Belair2002}. The drainage and subsequent bursting of bubbles at the surface of a liquid pool have therefore attracted some attention in the literature, mostly in the case of ``bare'' viscous bubbles\cite{Debregeas1998,Howell1999,Nguyen2013,Kocarkova2013} or in the presence of minute quantities of surface active agents\cite{Lhuissier2011,Modini2013} mimicking seawater composition.

The rupture scenario of bubbles floating at a liquid surface can be summarized as follows. Due to gravity or capillary suction, the liquid flows down in the bubble cap, leading to a thinner and thinner film. A hole eventually nucleates in the film and propagates. The rim surrounding the hole can destabilise into small droplets, which are ejected, or even fold and produce daughter bubbles by air entrapment\cite{Bird2010}. The question of when and where a bubble bursts is relevant to aerosol generation since the bubble lifetime sets the average cap thickness at bursting, and thus the aerosol size distribution, while the position of the nucleation point influences the dispersion of these aerosols in the surrounding atmosphere. 

Bubble drainage and rupture are \textit{a priori} coupled since a thinner film is expected to be more prone to bursting\cite{Rio2014}. For surfactant-free films, for example, Vrij \& Overbeek\cite{Vrij1968} showed that the typical time required for a van der Waals driven thickness instability to grow is proportional to the film thickness to the power of $5$, thus leading to the earlier rupture of thinner films. In the absence of surfactants, \textit{i.e.} in the case of ``bare'' bubbles, the liquid/air interfaces are stress-free and the film thickness $h$ at the apex is found to decay exponentially with time\cite{Debregeas1998,Nguyen2013,Kocarkova2013}. As expected from intuition, viscous silicone oil bubbles were observed to puncture spontanously at the apex, where the film is the thinnest\cite{Debregeas1998}.
%

The presence of surfactants, even in tiny amounts, alters the bubble thinning dynamics in a way that is not yet fully understood. On the one hand, interfacial stresses due to surface tension gradients or surface viscosity\cite{Bhamla2014} can arise and slow down the drainage dynamics. On the other hand, additionnal thinning mechanisms, such as marginal regeneration, can appear in the presence of surfactants. Marginal regeneration, as introduced by Mysels \textit{et al.}\cite{Mysels1959} consists in the creation of thin ``pinching'' zones in a film, where it connects to a meniscus. The origin of these pinching regions remains unclear: they were shown to arise from capillary suction in a film with a rigid boundary condition at the interfaces\cite{Aradian2001}, but are not expected in films with stress-free interfaces\cite{Howell2005}. It was also proposed that marginal regeneration may stem from a surface tension gradient between the film and the meniscus\cite{Nierstrasz1998, Nierstrasz1999}. Whatever its source, marginal regeneration contributes to the overall film thinning by the rising of thinner patches generated in the vicinity of the meniscus\cite{Mysels1959} and is thought to play an important role in the drainage of bubbles made from very diluted surfactant solutions\cite{Lhuissier2011,Modini2013}. Additionally, the observation that tap water bubbles preferentially puncture close to their foot -- \textit{i.e.} where the thinner patches are produced -- led Lhuissier \& Villermaux\cite{Lhuissier2011} to suggest that marginal regeneration may be at the origin of bubble bursting. Nevertheless, it is still unclear how the location of bursting depends on the concentration of surface-active impurities in water.

In this article, we investigate the influence of surfactant concentration on the drainage mechanism in surfactant-stabilised bubbles and its relation to film bursting. We perform experiments on bubbles made from TTAB (tetradecyl trimethylammonium bromide) solutions of various concentrations, both above and below the critical micellar concentration (cmc). The corresponding experimental setup is presented in Section \ref{sec:setup}. In Section \ref{sec:drainage}, we report our measurements on the drainage dynamics at the apex and interpret them using an extrapolation length $b$, which characterizes the flow profile resulting from the stress balance at the liquid/air interfaces. Section \ref{sec:bursting} contains our results on the position of the rupture nucleation point, which we put into perspective with the drainage measurements, before concluding in Section \ref{sec:conclusion}.
%
%
%
\section{Experimental setup} \label{sec:setup}
%
\subsection{Bubble generation}
%
\begin{figure}
\centering
\includegraphics[width=\linewidth]{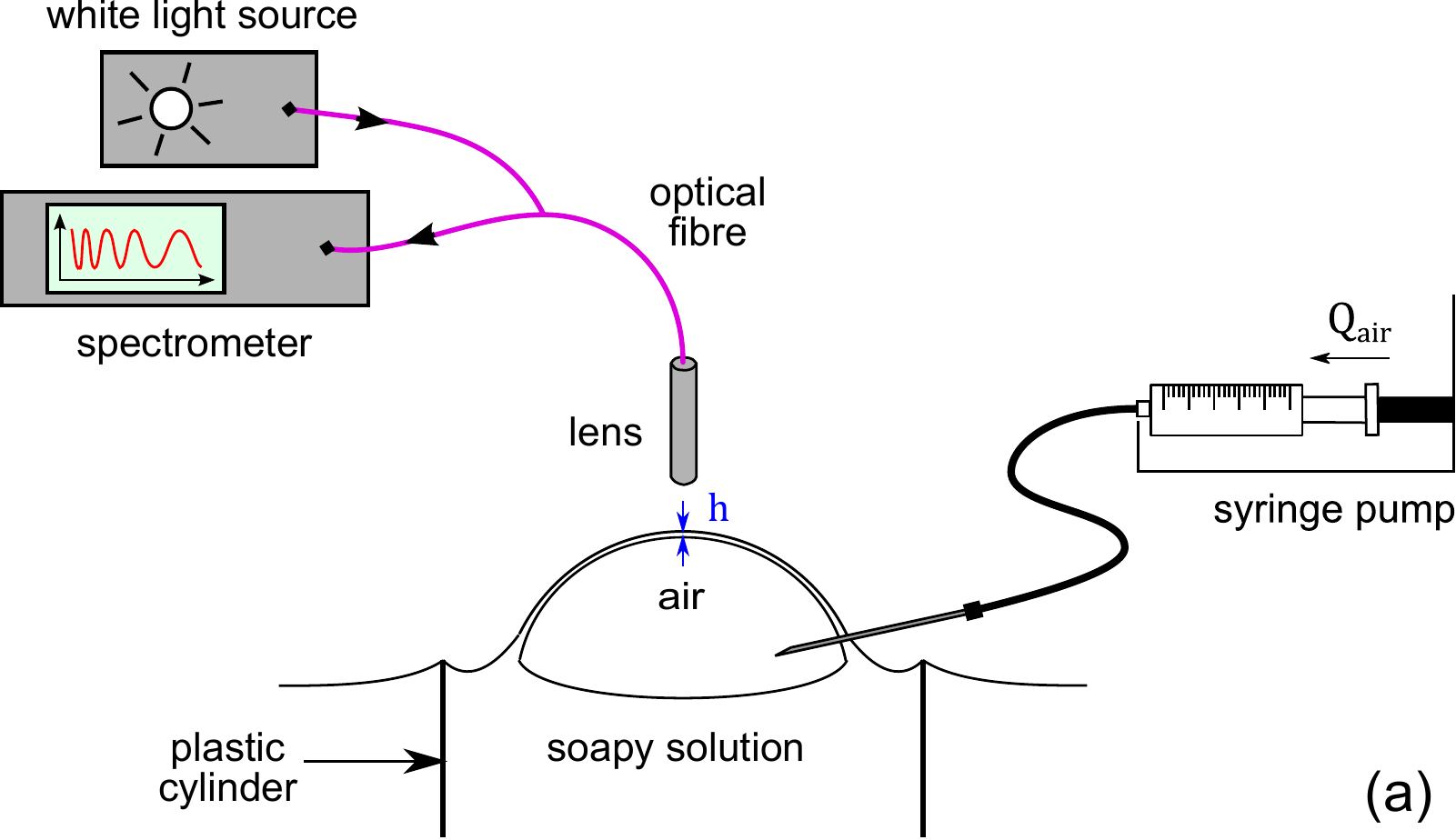} \smallskip\\
\includegraphics[width=\linewidth]{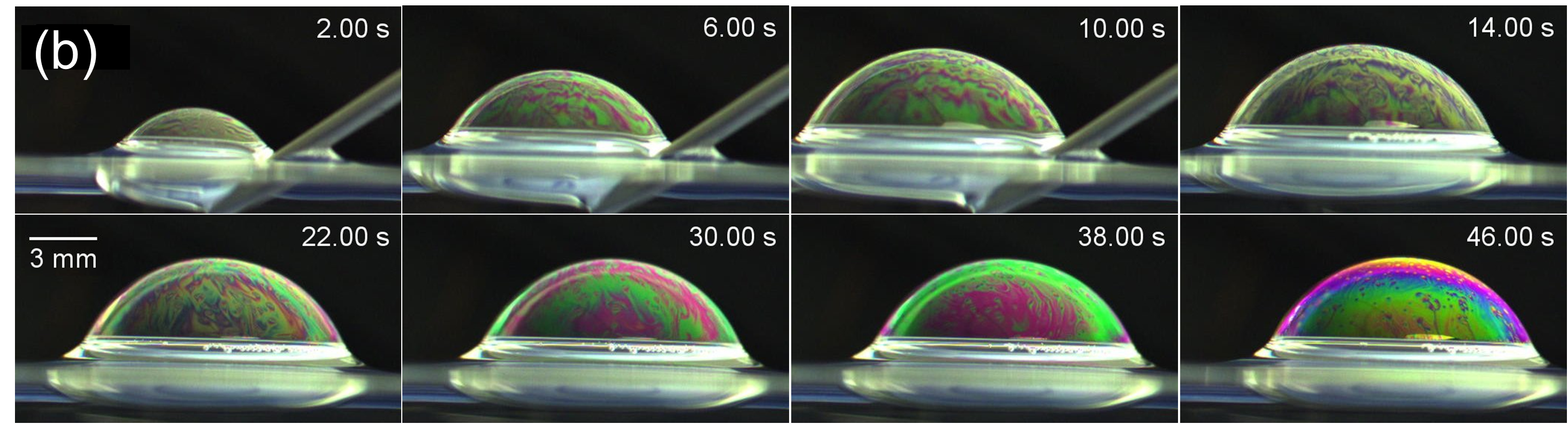}
\caption{(a) Scheme of the setup used to study air bubbles at the surface of a liquid pool. The bubble of volume $V$ is generated by injecting air just underneath the liquid surface with a controlled flow rate $\Qair$. Thickness measurements at the apex are performed with a white-light interferometric technique, while the bubble is kept immobile using a plastic cylinder. (b) Time evolution of a bubble stabilized by TTAB at $c=10~\cmc$, showing the color change due to bubble drainage. In this sequence, the bubble is blown at a flow rate $\Qair =4~\milli\liter\per\minute$ and has a final volume $V=0.8$ mL.
}
\label{fig:generation}
\end{figure}
The surfactant used in this study is TTAB (tetradecyl trimethylammonium bromide, purchased from Sigma-Aldrich), a cationic surfactant of critical micellar concentration $\cmc = 3.6~\milli\mole\per\liter$. Aqueous TTAB solutions of concentration $c$ ranging from $0.01$ and $10~\cmc$ are made with ultrapure water (resistivity $>18.2~\mega\ohm \cdot \centi\meter$) obtained from a Millipore Simplicity 185 device. The samples are prepared in plastic containers in order to avoid the electrostatically driven adsorption of TTAB onto glass surfaces. The solutions are sealed and used within one week of preparation.

The solution is poured into a petri dish of $5~\centi\meter$ in diameter and $5-10~\milli\meter$ in depth up to the brim. All petri dishes are rinsed with ethanol and then ultrapure water before use. As sketched in Figure \ref{fig:generation}a, a bubble is blown by injecting air just below the solution surface, through a needle (Terumo) of internal diameter $1.2~\milli\meter$. The air flow rate $\Qair$ and total injected volume $V$ are controlled using a syringe pump (KD Scientific). Once the target volume $V$ has been reached, air injection stops and the needle is either left in place or delicately removed. The bubble is then left to drain until it bursts.
%
\subsection{Bubble observation and thickness measurements}
%
From generation to bursting, a side-view of the bubble is recorded with a color camera (U-Eye), allowing bubble lifetime measurement as well as observation of the motions of the colored fringes that result from the interference between the successive reflexions of white light on both interfaces of the bubble cap. A typical image sequence corresponding to bubble generation and drainage is presented in Figure \ref{fig:generation}b for a $10~\cmc$ TTAB solution. The bubble bursting is recorded at frame rates ranging from 10 000 to 25 000 frames per second using a high-speed camera (Photron Fastcam SA3), allowing to locate the rupture nucleation point.

In order to obtain a more quantitative insight into bubble drainage, the film thickness at the apex is measured as a function of time using a white light interferometric technique, as sketched in Figure \ref{fig:generation}a. Polychromatic light emitted by a halogen lamp is carried by an optical fiber placed vertically above the bubble and focused onto the bubble top by a lens. The reflected light spectrum is collected by the same optical fiber and analysed by a spectrometer (USB 400 Ocean Optics), of bandwidth $350-1000~\nano\meter$. The film thickness at the apex is obtained by fitting the spectrum with help of the NanoCalc thin film metrology software (Mikropack, Ocean Optics).

A good spectrum contrast is observed only if the bubble surface is exactly perpendicular to the optical fiber. Thickness measurements thus require accurate control of the bubble position. This is achieved by blowing the bubble in the area enclosed by a plastic cylinder (see Figure \ref{fig:generation}a), on the edge of which the bubble meniscus is pinned. Fine position adjustments can be made thanks to a $x-y$ horizontal translation plate on which the petri dish is set.
%
%
\section{Bubble drainage} \label{sec:drainage}
%
\subsection{Experimental results and comparison to existing models}
%
\begin{figure}
\centering
\includegraphics[width=\linewidth]{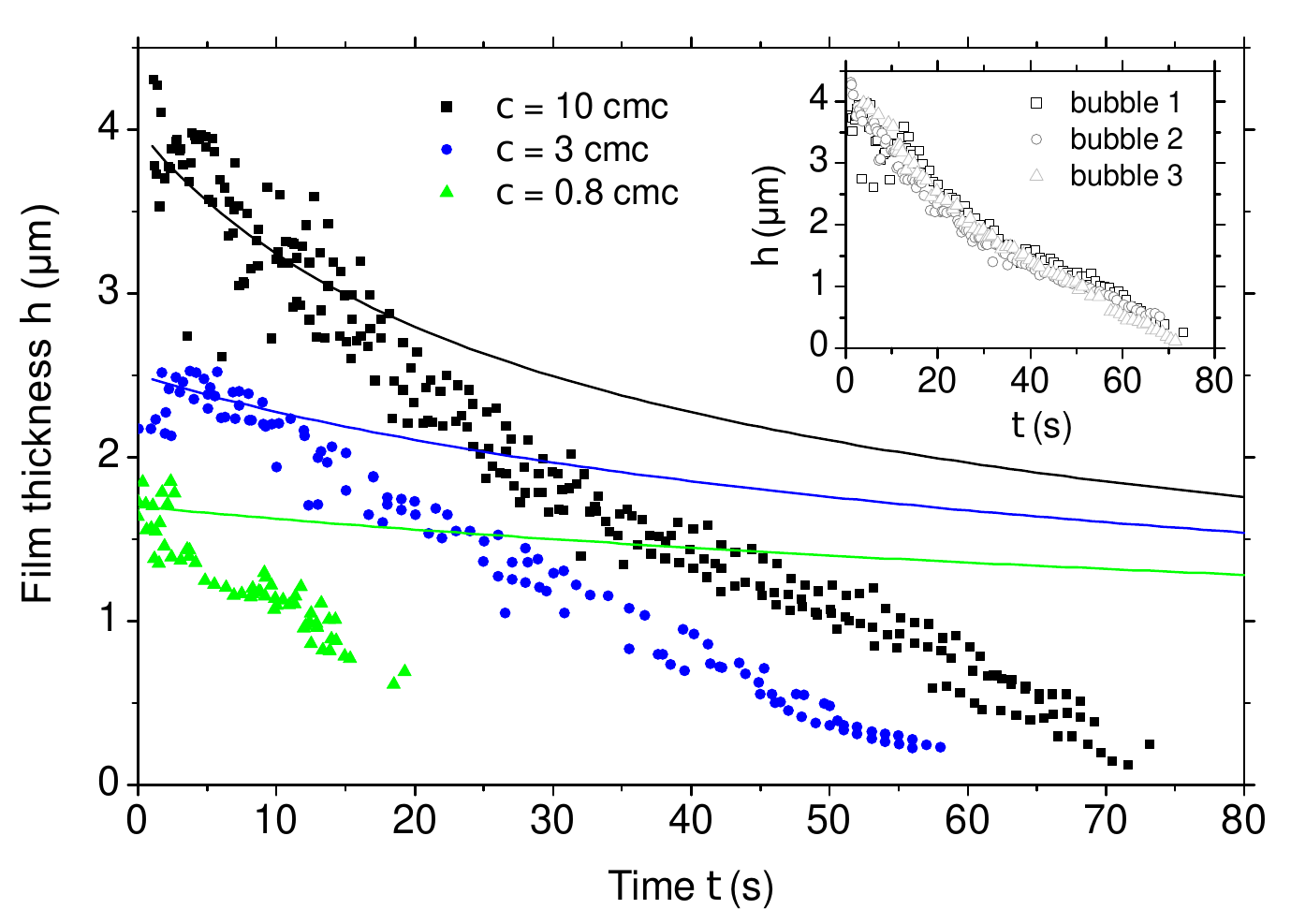}
\caption{Time evolution of the film thickness $h$ at the bubble apex for various TTAB concentrations (colors) and a fixed air flowrate $\Qair =4~\milli\liter\per\minute$. For each concentration, we superimpose the data corresponding to three different bubbles (symbols) and compare the drainage curve to the prediction in the limit of rigid liquid/air interfaces (Equation \eqref{eq:drainage_bulle_rigide}, solid lines). The inset shows the reproducibility of the drainage curve for three different bubbles generated in the same conditions from a $10~\cmc$ TTAB solution.}
\label{fig:raw_data_drainage}
\end{figure}
%
%
The film thickness $h$ at the apex of $V=1~\milli\liter$ bubbles generated at a flow rate $\Qair =4~\milli\liter\per\minute$ is measured as a function of time for different TTAB concentrations $c=0.8$, $3$ and $10~\cmc$. The thickness measurement starts at the beginning of the free drainage phase, once the bubble has reached its final volume.

The insert in Figure \ref{fig:raw_data_drainage} shows the drainage curve at the apex of three bubbles generated in the same conditions from a $10~\cmc$ TTAB solution. The initial thickness and the drainage dynamics turn out to be reproducible. These data sets are represented again in Figure \ref{fig:raw_data_drainage} along with the drainage curves obtained for lower surfactant concentrations ($c=0.8$ and $3~\cmc$). In each case, the data corresponding to three different bubbles are superimposed (symbols in Figure \ref{fig:raw_data_drainage}). Measurements for surfactant concentrations below $0.8~\cmc$ could not be achieved because the bubbles burst immediately upon contacting the edge of the plastic cylinder holding them in place. 

\begin{figure}
\centering
\includegraphics[width=\linewidth]{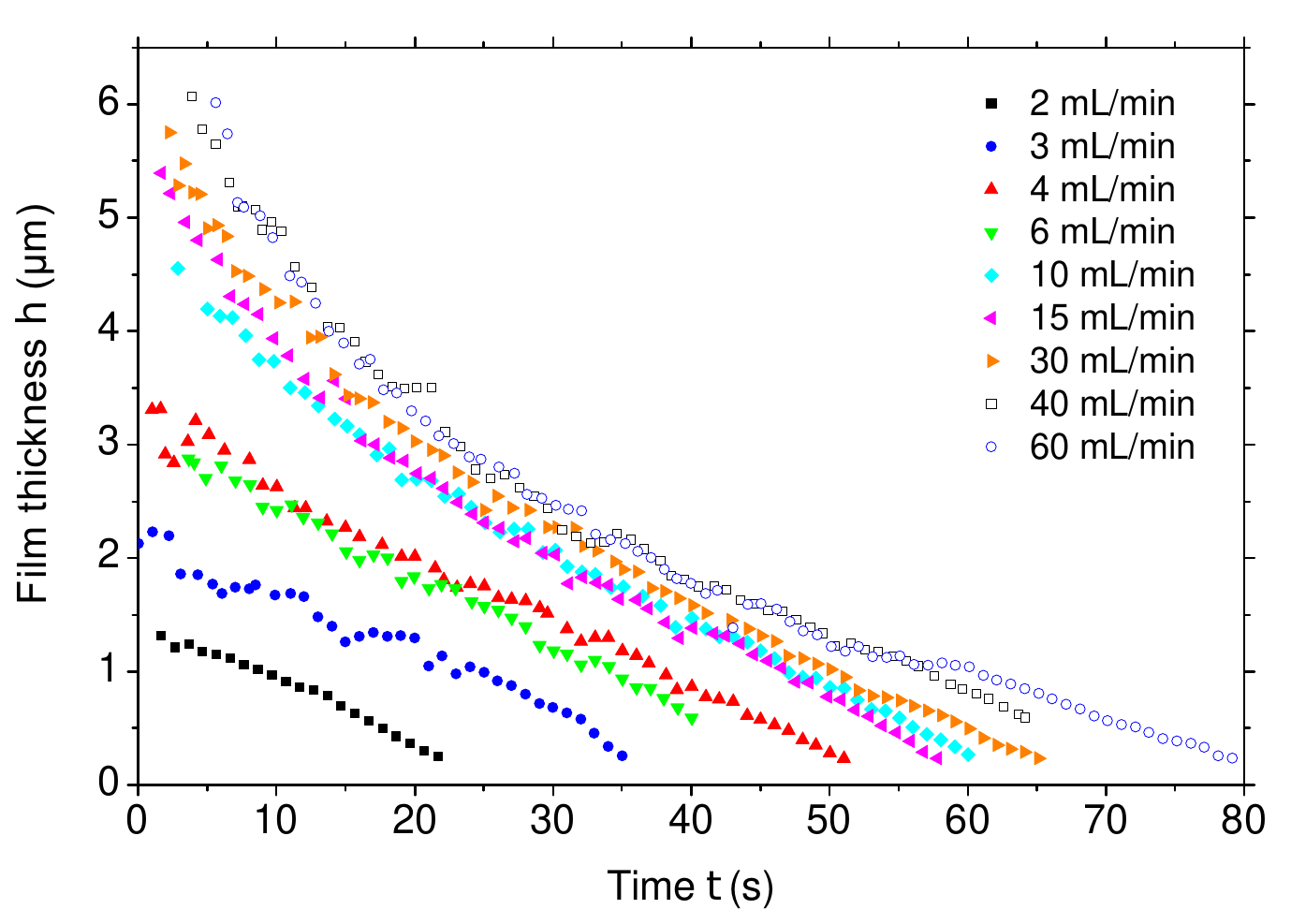}
\caption{Time evolution of the film thickness $h$ at the bubble apex for a fixed TTAB concentration ($c=10~\cmc$) and various air flowrates $\Qair$ plotted in log-linear scale. For each flowrate, the initial time $t=0$ corresponds to the end of bubble inflation, when the bubble has reached its final volume.}
\label{fig:raw_data_scan_Qair}
\end{figure}

For a fixed concentration $c=10~\cmc$, we also explore the influence of the air flowrate $\Qair$ at which the bubble is inflated. Figure \ref{fig:raw_data_scan_Qair} shows the drainage curves measured at the apex of $1~\milli\liter$ bubbles, for $\Qair$ ranging from $2~\milli\liter\per\minute$ (corresponding to a generation time of $30~\second$) to $60~\milli\liter\per\minute$ (corresponding to a generation time of $1~\second$). We stress that the time represented in Figure \ref{fig:raw_data_scan_Qair} is the real time minus the bubble generation time, so that $t=0$ corresponds to the beginning of bubble free drainage, \textit{i.e.} to the end of bubble inflation. \medskip \\
\indent In order to discuss the drainage dynamics observed in Figure \ref{fig:raw_data_drainage}, let us compute the Bond number $\Bo \equiv \rho g R^2/\gamma$, which compares the hydrostatic pressure scale $\rho g R$ (with $\rho$ the liquid density and $g$ the gravitational acceleration) to the capillary pressure scale $\gamma / R$ (with $\gamma$ the surface tension) for a bubble of radius $R$. Introducing the capillary length $\ell_c \equiv \sqrt{\gamma/\rho g}$, the Bond number can be simply rewritten as $\Bo = (R /\ell_c)^2$. In our experiments, the radius of $1~\milli\liter$ bubbles is $R \approx 9~\milli\meter$, while the surface tension of TTAB solutions \cite{Bergeron1997} ranges from $38.5~\milli\newton\per\meter$ for $c=3$ and $10~\cmc$ to $41.8~\milli\newton\per\meter$ for $c=0.8~\cmc$, yielding Bond numbers of $\Bo = 21$ and $\Bo = 19$ respectively. In all cases, we have $\Bo \gg 1$, meaning that bubble drainage is mainly driven by gravity.

In the presence of surfactants, it seems natural to compare experimental data to a gravitational drainage model with large interfacial stresses. In this limit, the velocity at the interfaces tends to zero, leading to a rigid boundary condition at the interfaces. The film thickness $h$ at the apex then follows an algebraic decay with time $t$, of the form \cite{Lhuissier2011,Bhamla2014}
\begin{equation}
h(t) = \frac{h_0}{\sqrt{1 + t/3\Tr}} \qquad \text{with} \qquad \Tr = \frac{\eta R}{\rho g h_0^2},
\label{eq:drainage_bulle_rigide}
\end{equation}
where $h_0$ is the initial film thickness and $\eta$ the liquid viscosity. This model is represented by the solid lines in Figure \ref{fig:raw_data_drainage}. The experimentally observed drainage dynamics is however significantly faster than predicted by Equation \eqref{eq:drainage_bulle_rigide} or by more general functions of the same form, as developed in the ESI (section 2). Consequently, we now examine the opposite limit of stress-free interfaces.

In this limit and for the gravitational drainage of a bubble of radius $R$, Ko{\v{c}}{\'a}rkov{\'a} \textit{et al.} predict an exponential decay \cite{Debregeas1998,Nguyen2013} and propose the expression \cite{Kocarkova2013}
\begin{equation}
h(t) = h_0 \, \exp (-at/\Tsf) \qquad \text{with} \qquad \Tsf = \frac{\eta}{2R\rho g}.
\label{eq:drainage_stress-free}
\end{equation}
They showed both experimentally and theoretically that the thinning rate $a$ in Equation \eqref{eq:drainage_stress-free} is a decreasing function of the Bond number $\Bo$ and tends to $a \approx 0.1$ in the limit of $\Bo \gg 1$ corresponding to our experimental conditions. The characteristic time $\Tsf = \eta / 2R \rho g$ is of the order of $10~\micro\second$ for our aqueous TTAB solutions, hence a characteristic drainage time $\Tsf /a$ of the order of $0.1~\milli\second$ in the limit of large Bond numbers. We observe experimentally that drainage typically proceeds over tens of seconds, which is clearly incompatible with the hypothesis of stress-free interfaces. In the following, we thus propose a simple model for graviational drainage accounting for intermediate -- \textit{i.e.} partially rigid -- boundary conditions at the bubble liquid/air interfaces.

Note that, despite the presence of marginal regeneration features at all the surfactant concentrations tested, the model proposed by Lhuissier \& Villermaux\cite{Lhuissier2011} for marginal regeneration driven drainage of bubbles fails to account for our experimental data, as shown in the ESI (section 1).
%
\subsection{Drainage model with partially rigid interfaces}
%
\begin{figure}
\centering
\includegraphics[width=\linewidth]{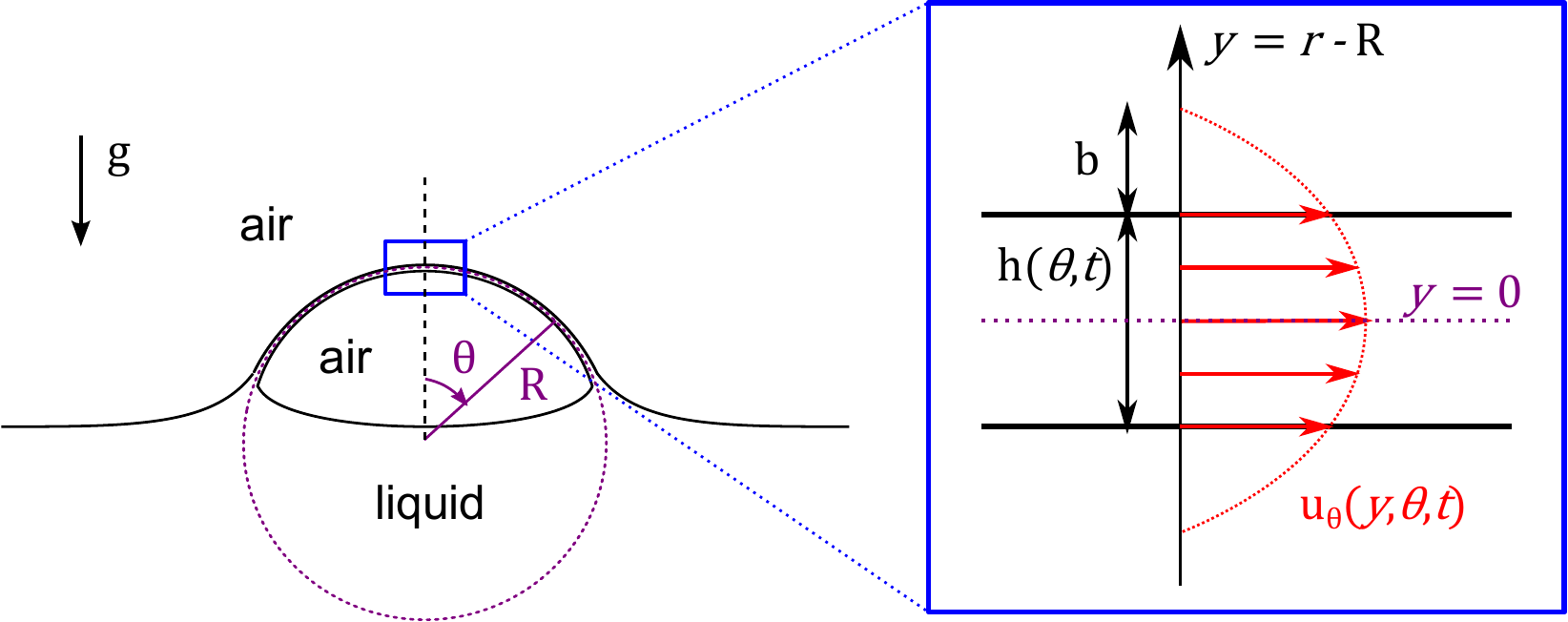}
\caption{Scheme of the bubble with spherical coordinates used in the model. The enlargment on the right introduces the notations used to describe the liquid flow in the film. In particular, we define the extrapolation length $b \geq 0$ at which the parabolic velocity profile $\uth (y,\theta, t)$ becomes zero.}
\label{fig:notation_drainage_bulle}
\end{figure}
We seek to describe the gravitational drainage of a hemispherical bubble of radius $R$ greater than the capillary length $\ell_c$, as sketched in Figure \ref{fig:notation_drainage_bulle}. Computing the actual surface stress at the interfaces of a surfactant-stabilised bubble is not an easy task, since it results from viscous surface stresses related to interfacial velocity, and from surface tension gradients, which in turn depend on the local surfactant concentration  \cite{Edwards1991}. Instead, we propose a more geometrical description of the flow profile in the bubble cap, based on the introduction of an extrapolation length $b$, which is approximated as being independent of time and position. As represented on the right panel in Figure \ref{fig:notation_drainage_bulle}, $b$ is defined as the distance from the interface where the tangential velocity field $\uth$ -- which has the classical parabolic shape -- goes to zero.

The bubble radius $R \approx 9~\milli\meter$ being much larger than the typical film thickness, we use the lubrication approximation. The tangential velocity field $\uth (y,\theta, t)$ is then governed by Stokes' equation \cite{Bhamla2014}
\begin{equation}
\eta \, \partial_{yy} \uth = -\rho g \sin \theta,
\label{eq:Stokes_sphérique}
\end{equation}
where the coordinate $y=r-R$ is defined by translating the usual radial coordinate $r$ so that the liquid film is located in the interval $y \in [-h/2, h/2]$. Integrating Stokes' equation \eqref{eq:Stokes_sphérique} with the symmetry condition $\partial_y \uth = 0$ and the definition of the extrapolation length $\uth (y=h/2+b) = 0$, we can compute the tangential velocity profile in the film
\begin{equation}
\uth = - \frac{\rho g}{2\eta} \sin \theta \left[ y^2 - \left( \frac{h}{2} + b \right)^2 \right].
\label{eq:profil_vitesse_bulle}
\end{equation}
The time evolution of the film thickness $h$ is then given by the mass conservation in the film, which reads in the lubrication approximation \cite{Scheid2012_antibubble} 
\begin{equation}
\partial_t h + \frac{1}{R \sin \theta} \partial_{\theta} \left( \sin \theta \, h \uthbar \right) = 0,
\label{eq:conservation_masse_bulle}
\end{equation}
where $\uthbar$ is the cross-sectionally averaged velocity at angle $\theta$
\begin{equation}
\uthbar (\theta, t) \equiv \frac{1}{h} \int_{-h/2}^{h/2} \uth (y, \theta, t) \, \dd y%
= - \frac{\rho g}{2 \eta} \sin \theta \left[ \frac{h^2}{12} - \left( \frac{h}{2} + b \right)^2 \right].
\label{eq:vitesse_moyenne_bulle}
\end{equation}
Under the assumption that the extrapolation length $b$ does not depend on the angle $\theta$, the mass conservation (Equation \eqref{eq:conservation_masse_bulle}) finally becomes
\begin{multline}
\partial_t h - \frac{\rho g}{2 \eta R} \left\lbrace 2 \cos\theta \left[ \frac{h^3}{12} - h\left( \frac{h}{2} + b \right)^2 \right] \right. \\
- \left. \sin \theta \, \partial_{\theta} h \left( \frac{h^2}{2} + 2hb + b^2 \right) \right\rbrace = 0. 
\label{eq:cons_masse_bulle_développée}
\end{multline} 
%
Out of the two terms between the braces, only the left one remains when Equation \eqref{eq:cons_masse_bulle_développée} is evaluated at the bubble apex ($\theta =0$). Consequently, the spatial derivative disappears from Equation \eqref{eq:cons_masse_bulle_développée}, leading to the ordinary differential equation 
\begin{equation}
\partial_t h + \frac{\rho g}{\eta R} \left( \frac{h^3}{6} + b h^2 + b^2 h \right) = 0.
\label{eq:cons_masse_bulle_apex}
\end{equation}
For $b \ll h/2$, we recover the equation governing the bubble drainage under the assumption of rigid liquid/air interfaces, whose solution is given by Equation \eqref{eq:drainage_bulle_rigide}, and the flow in the film is a classical Poiseuille flow with zero velocity at the interfaces. In the opposite limit of $b \gg h/2$, Equation \eqref{eq:cons_masse_bulle_apex} becomes linear -- provided that $b$ is constant in time -- and yields an exponential decay of the film thickness
\begin{equation}
h (t) = h_0 \exp \left( - t/ T \right) \qquad \text{with} \qquad T \equiv \frac{\eta R}{\rho g b^2}.
\label{eq:drainage_grandb}
\end{equation}
The flow in the film is then getting closer to a plug flow. Finally, when $b \sim h/2$, all terms must be kept in Equation \eqref{eq:cons_masse_bulle_apex}, which has to be numerically integrated.
%
\subsection{Comparison to experimental data} \label{sec:comp_exp_drainage}
%
\begin{figure}
\centering
\includegraphics[width=\linewidth]{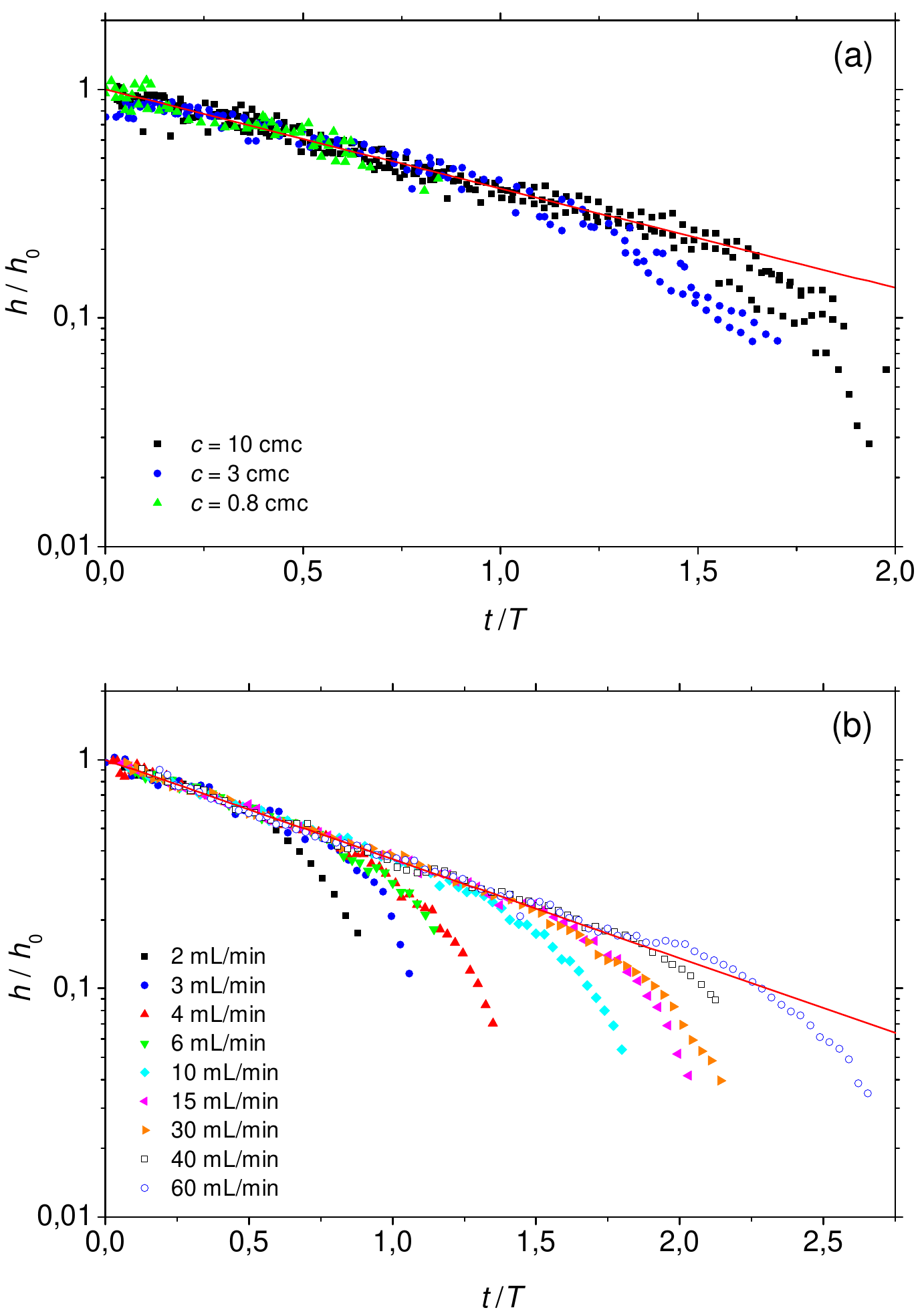}
\caption{Rescaled drainage curves at the bubble apex for (a) various TTAB concentrations $c$ and a fixed air flowrate $\Qair = 4~\milli\liter\per\minute$ and (b) various air flowrates $\Qair$ and a fixed TTAB concentration $c=10~\cmc$. The thickness and time are respectively normalized by the initial thickness $h_0$ and the characteristic time $T$ obtained by fitting the data with Equation \eqref{eq:drainage_grandb} (solid red line).}
\label{fig:rescaled_data_all}
\end{figure}
\begin{figure}
\centering
\includegraphics[width=\linewidth]{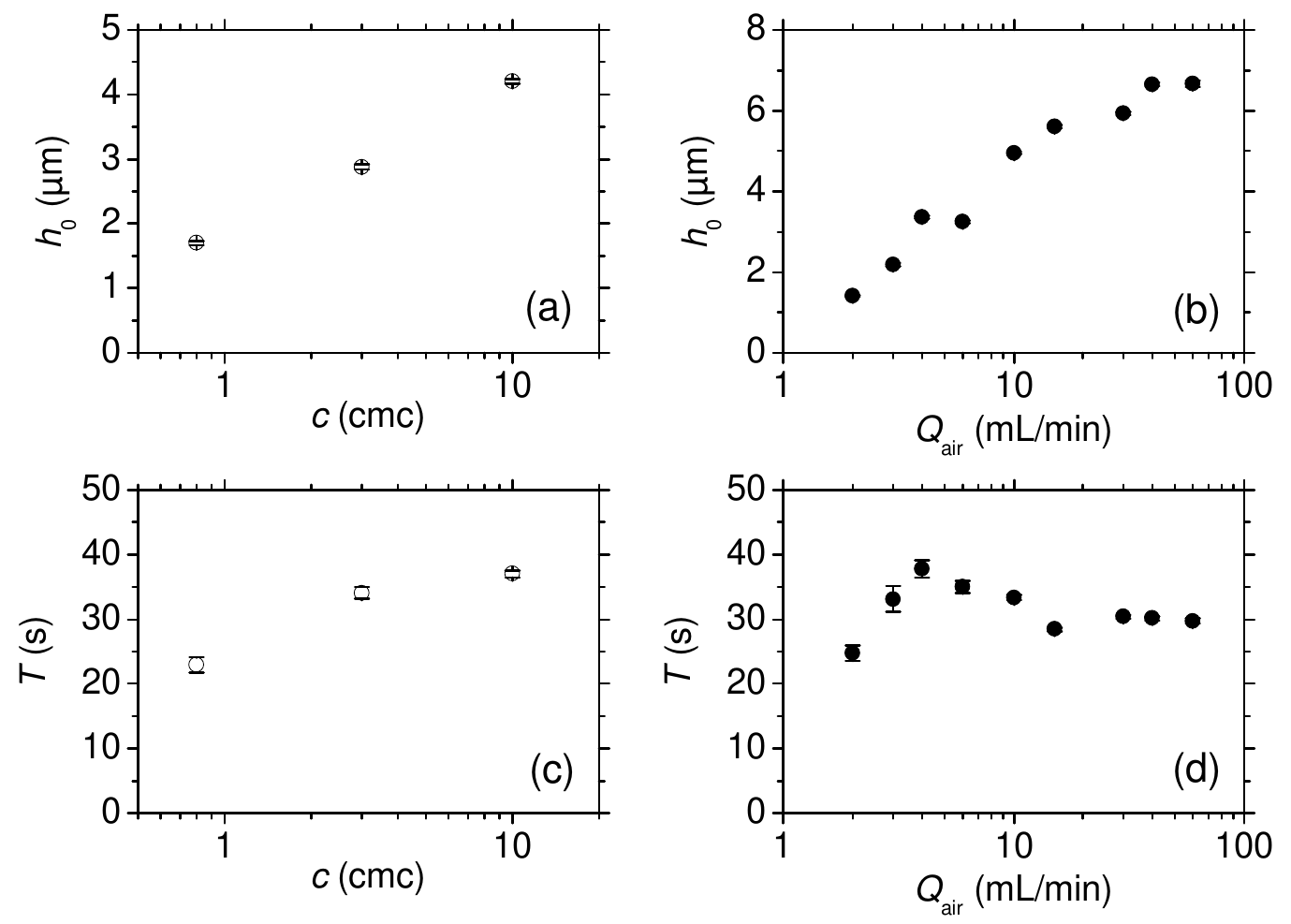}
\caption{Values of the initial thickness $h_0$ and the characteristic timescale $T$ obtained by fitting the experimental data with Equation \eqref{eq:drainage_grandb}, as a function of the surfactant concentration $c$ (a and c, respectively) and the air flowrate $\Qair$ (b and d, respectively).}
\label{fig:fit_param_all}
\end{figure}
\begin{figure}
\centering
\includegraphics[width=\linewidth]{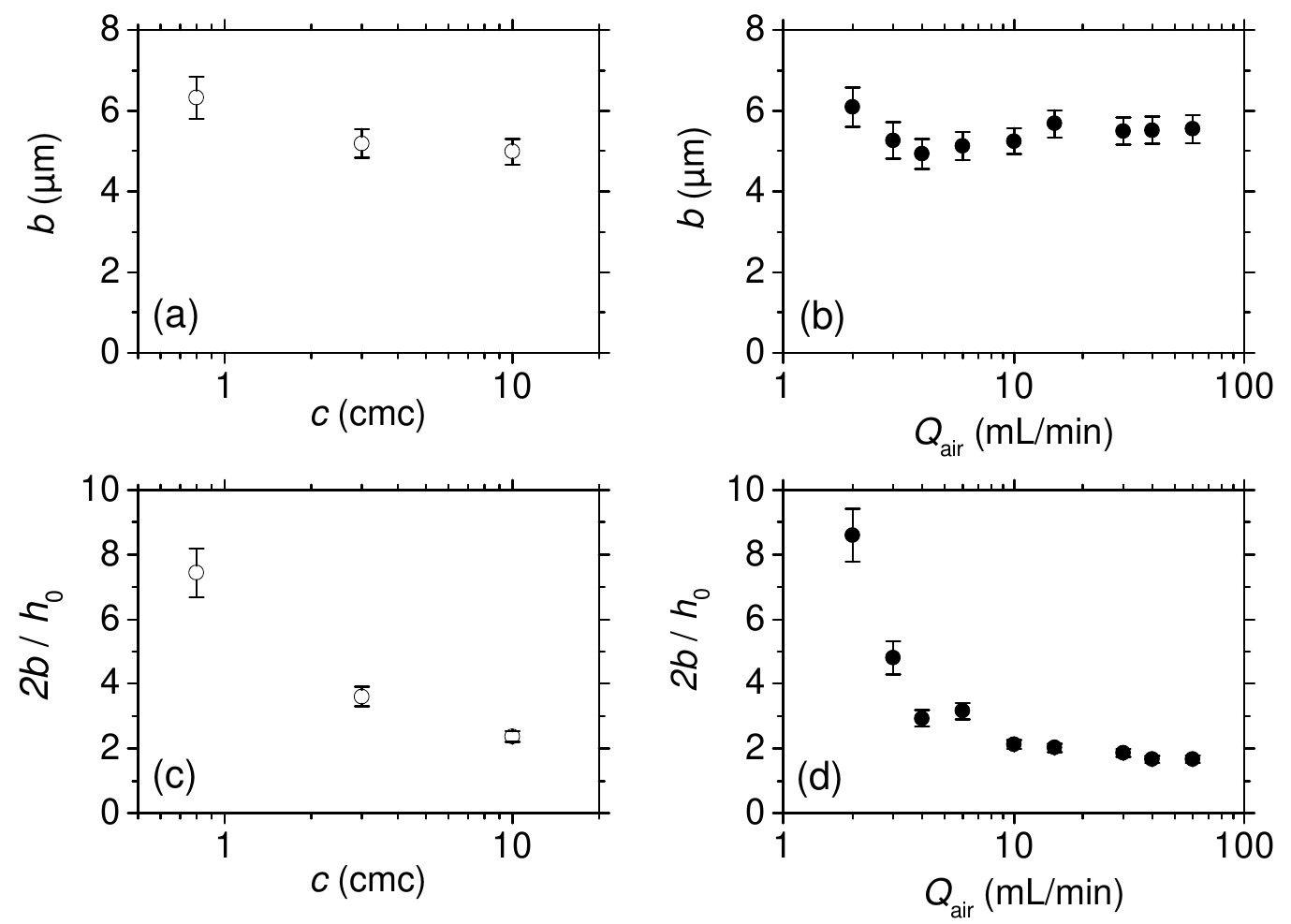}
\caption{Values of the extrapolation length $b$ and ratio $2b/h_0$ deduced from the characteristic timescale $T$ and initial thickness $h_0$ (Figure \ref{fig:fit_param_all}), as a function of the surfactant concentration $c$ (a and c, respectively) and the air flowrate $\Qair$ (b and d, respectively).}
\label{fig:extrapolation_length}
\end{figure}
In the following, we make the \textit{ad-hoc} assumption that $b \gg h/2$, meaning that the flow profile in the film is closer to a plug flow than to a Poiseuille flow. The experimental data presented in Figure \ref{fig:raw_data_drainage} are fitted with Equation \eqref{eq:drainage_grandb}, using the initial thickness $h_0$ and the characteristic time $T$ as adjustable parameters. For each TTAB concentration, the fitted values of $h_0$ and $T$ are reported in Figure \ref{fig:fit_param_all} (a and c) and used in Figure \ref{fig:rescaled_data_all}a to rescale the experimental data. The same is done for the experimental data from Figure \ref{fig:raw_data_scan_Qair}, obtained for various air flowrates $\Qair$, and the rescaled data are presented in Figure \ref{fig:rescaled_data_all}b. The resulting values of $h_0$ and $T$ are plotted in Figure \ref{fig:fit_param_all}b and d as functions of the air flowrate $\Qair$. In both Figures \ref{fig:rescaled_data_all}a and b, the prediction of Equation \eqref{eq:drainage_grandb} (solid red line) is then in good agreement with experimental data, except for the end of the drainage dynamics, where the thickness decreases faster than expected from the model.

For a fixed surfactant concentration, the initial film thickness $h_0$ is found to rise by a factor $6$ when increasing the flowrate from $2$ to $60~\milli\liter\per\minute$ (Figure \ref{fig:fit_param_all}b). This variation of $h_0$ can be understood qualitatively from the fact that the bubble already drains during its generation. Thus, the smaller the flowrate, the longer the generation time and the thinner the film when the bubble reaches its final volume ($t=0$ in Figures \ref{fig:raw_data_scan_Qair} and \ref{fig:rescaled_data_all}b). By analogy with the generation of flat vertical films, whose thickness is known to increase with the pulling velocity \cite{Mysels1959}, we can also anticipate that the thickness at the apex of bubbles will increase with the air flow rate, which controls the rate of production of fresh interface during bubble generation. Both phenomena probably contribute to the observed dependency of $h_0$ with $\Qair$.

For a fixed air flowrate, Figure \ref{fig:fit_param_all}a shows that the initial film thickness $h_0$ at the apex rises when the TTAB concentration $c$ is increased, going from $h_0 \approx 1.7~\micro\meter$ for $c=0.8~\cmc$ up to $h_0 \approx 4~\micro\meter$ for $c=10~\cmc$. This variation is likely due to an increased tangential stress at interfaces during bubble generation at higher TTAB concentrations. For instance, it has been shown that the thickness of flat vertical films not only increases with the entrainement speed, but also with the surface viscous stress \cite{Scheid2010_LLD} and the surface tension gradient \cite{Champougny2015} at the liquid/air interfaces, for a given pulling velocity.

Figures \ref{fig:fit_param_all}c and d finally indicate that the characteristic timescale $T$ for drainage depends only weakly on both the surfactant concentration $c$ and the air flowrate $\Qair$, and remains of the order of  $30~\second$. The extrapolation length $b$ can be deduced from the timescale $T$ given in Figures \ref{fig:fit_param_all}c and d through the relation
\begin{equation}
b = \sqrt{\frac{\eta R}{\rho g T}},
\label{eq:b_vs_échelle_temps}
\end{equation}
where the bubble radius is $R \approx 9~\milli\meter$. The resulting values of $b$ as a function of the TTAB concentration and the air flowrate are respectively displayed in Figures \ref{fig:extrapolation_length}a and b. The initial flow profile in the film is characterised by the ratio $2b/h_0$, which we plot as a function of the surfactant concentration and flowrate $\Qair$ in Figures \ref{fig:extrapolation_length}c and d, respectively. The hypothesis of a dominant plug flow ($2b/h_0 \gg 1$) is well justified for $c=0.8~\cmc$ and small flowrates, while it seems to be just validated for the most concentrated solutions and flowrates $\Qair \geqslant 4~\milli\liter\per\minute$, where $2b /h_0$ is of order $2-3$. The decrease of $2b/h_0$ with increasing the TTAB concentration and flowrate corresponds to a decrease of the plug flow contribution to the flow in the film and thus to an increased tangential stress at the liquid/air interfaces. 

It can be noted from Figure \ref{fig:extrapolation_length} that the extrapolation length, of the order of $6~\micro\meter$, does not change much with surfactant concentration and air flowrate, and that the variations of $2b /h_0$ thus essentially stems from the contribution of $h_0$. A comparable value for the extrapolation length had been obtained in the work by Saulnier \textit{et al.} on vertical $\mathrm{C}_{12}\mathrm{E}_6$-stabilized films studied during their generation \cite{Saulnier2014}. Thickness measurements of PEO/SDS-stabilized films were also interpreted by Adelizzi \& Troian \cite{Adelizzi2004} using extrapolation lengths of the order of a few microns and Berg \textit{et al.} \cite{Berg2005} found extrapolation lengths about one order of magnitude smaller for SDS-stabilized vertical films.
%
\section{Bubble bursting} \label{sec:bursting}
%
We now turn to bubble bursting, which we characterise by measuring the position of the rupture nucleation point and the bubble lifetime. In this section, we study bubbles of fixed volume $V = 0.8~\milli\liter$, corresponding to a radius $R \approx 8~\milli\meter$, and the generation flow rate remains equal to $\Qair = 4~\milli\liter\per\minute$. Note that the plastic cylinder used to hold the bubble in place for drainage measurements is removed.
%
\subsection{Position of the rupture nucleation point} \label{sec:bursting_position}
%
\begin{figure}
\centering
\includegraphics[width=5cm]{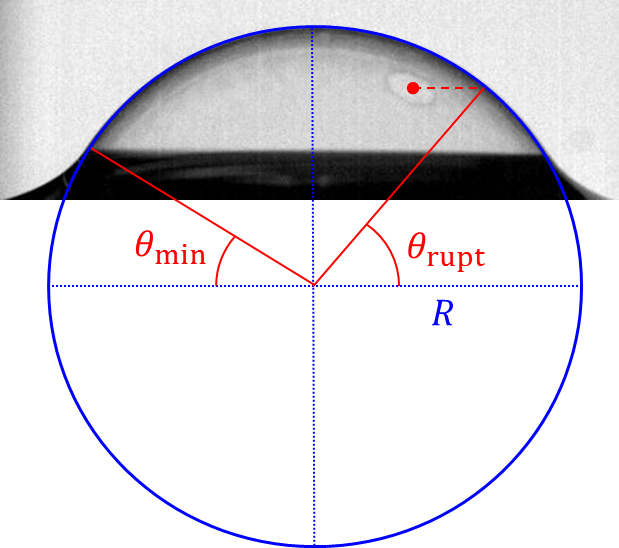}
\caption{Definition of the rupture angle $\thrupt$, which characterizes the position of the rupture nucleation point, and the angle $\thmin$, corresponding to the limit between the meniscus (in black) and the thin bubble cap.}
\label{fig:mesure_angle_rupture}
\end{figure}
\begin{figure}
\centering
\includegraphics[width=\linewidth]{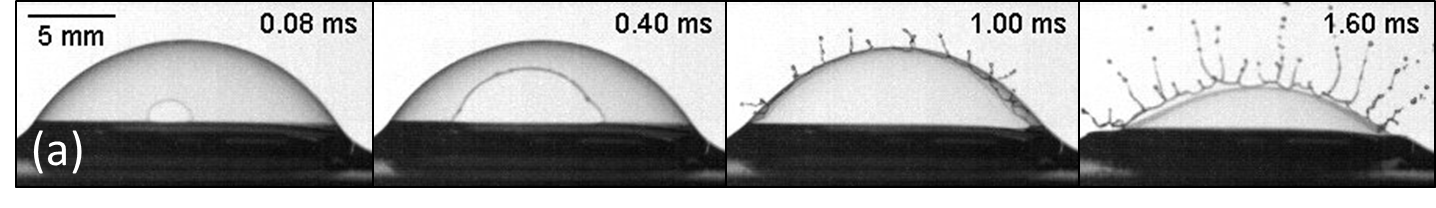}
\vspace{0.1cm} \\
\includegraphics[width=\linewidth]{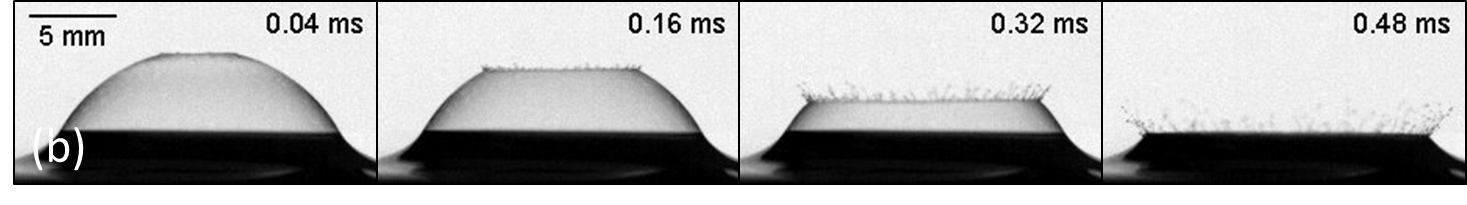}
\caption{Image sequences recorded with a high-speed camera at 25 000 frames per second, showing typical bursting events for TTAB concentrations of (a) $c=0.1~\cmc$ and (b) $c=10~\cmc$. In both cases, the initial time is defined as the time when the hole nucleation is observed and the corresponding error is $\pm 0.04~\milli\second$.}
\label{fig:sequences_rupture}
\end{figure}
Since the bubble has a (hemi)spherical symmetry, we characterise the position of the rupture nucleation point by its latitude $\thrupt$, as shown in Figure \ref{fig:mesure_angle_rupture}. Using the image analysis sofware ImageJ, the bubble cap is fitted by a circle, allowing us to define $\thrupt$ as the angle between the projection of the nucleation point onto the circle and the horizontal direction. Similarly, we also introduce the angle $\thmin$ (where ``min'' stands for ``minimum'') corresponding to the limit between the meniscus (in black) and the thin film. 

Because of the bubble symmetry with respect to the vertical axis, the rupture angle $\thrupt$ varies in the interval $[\thmin, \pi /2]$. In the following, the position of the rupture nucleation point will thus be characterised by the reduced rupture angle
\begin{equation}
\Theta = \frac{\thrupt - \thmin}{\pi/2 - \thmin},
\label{eq:angle_rupture_réduit}
\end{equation}
which ranges from $0$, when the hole nucleates at the foot of the bubble, to $1$, when the bursting is spontaneously initiated at the apex. \bigskip \\
The reduced rupture angle $\Theta$ is measured systematically as a function of TTAB concentration, ranging from $0.01$ to $10~\cmc$. As shown in Figure \ref{fig:sequences_rupture}, the angle $\Theta$ is observed to span the whole range of possible behaviours, from $\Theta = 0$ (Figure \ref{fig:sequences_rupture}a) to $\Theta =1$ (Figure \ref{fig:sequences_rupture}b). Various experimental conditions are tested: (i) leaving the needle in place or (ii) removing it with care after bubble blowing and (iii) enclosing the whole setup in a sealed glove box.

\begin{figure}
\centering
\includegraphics[width=\linewidth]{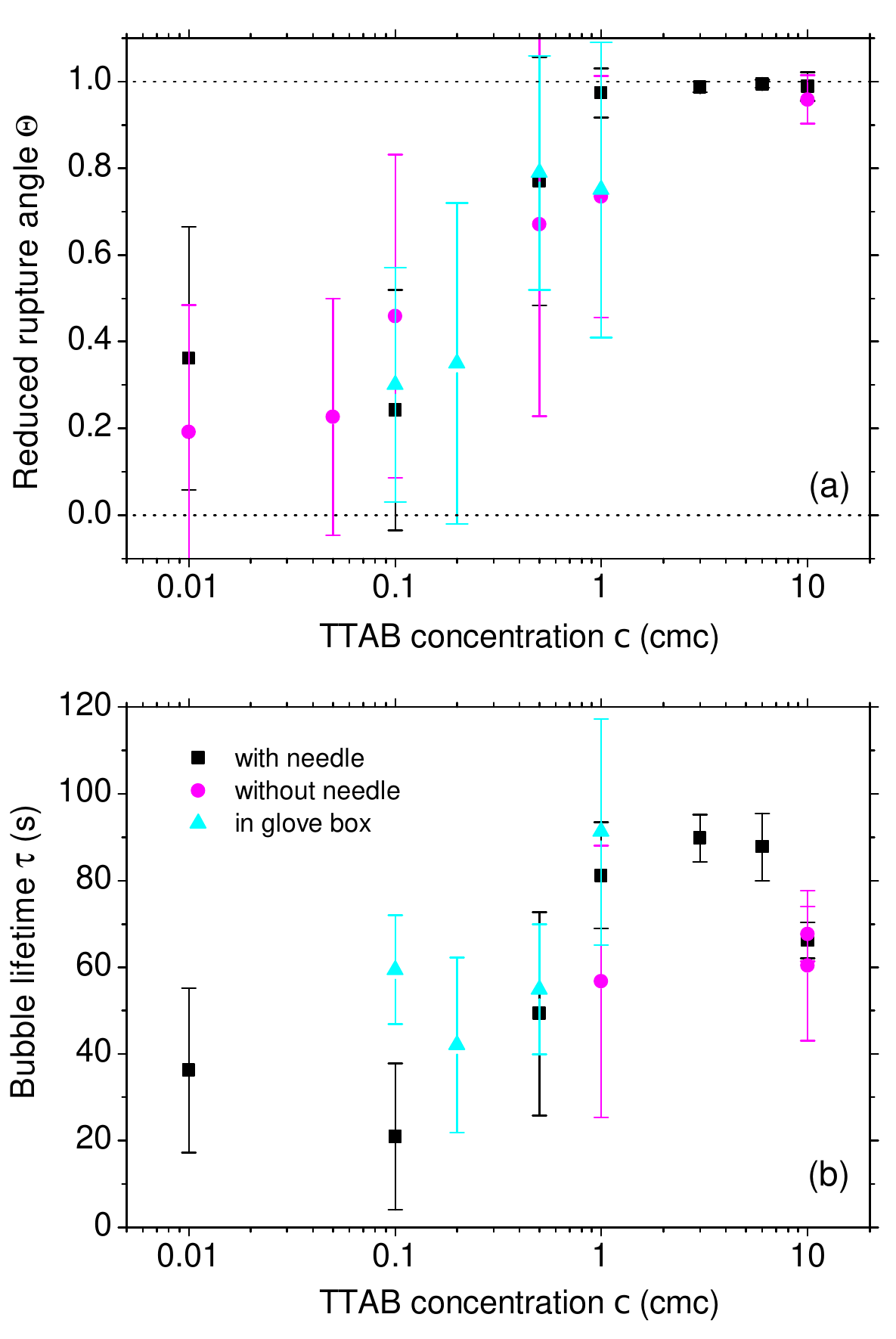}
\caption{(a) The reduced rupture angle $\Theta$, characterising the position of the rupture nucleation point (see Equation \eqref{eq:angle_rupture_réduit}) and (b) the lifetime $\tau$ of bubbles, counted from the beginning of bubble generation, are plotted as functions of TTAB concentration $c$. The different symbols represent different experimental conditions.}
\label{fig:moyennes_rupture}
\end{figure}

The results obtained in these different situations are displayed in Figure \ref{fig:moyennes_rupture}a, where each point is an average over 4 to 20 bubbles and the error bar stands for the standard deviation. For TTAB concentrations above the cmc, the position of the rupture nucleation point turns out to be reproducible and is located at the bubble apex as displayed in the image sequence in Figure \ref{fig:sequences_rupture}b. Below the cmc, however, the dispersion of data points becomes quite large and the position of the rupture nucleation point is no longer reproducible. On average, the reduced rupture angle $\Theta$ decreases when $c$ decreases, showing that more and more rupture events are observed close to the bubble foot ($\Theta = 0$) when the surfactant concentration diminishes. Finally, note that no significant difference was found between the data sets corresponding to the different experimental conditions mentioned above (with needle, without needle, experiments in glove box). 
%
%
\subsection{Bubble lifetime}
The bubble lifetime $\tau$ is defined as the time lapse between the beginning of bubble blowing and the bursting. In the following, we only take into account the bubbles that burst after the end of generation, corresponding to a minimal lifetime of $12~\second$ for $0.8~\milli\liter$ bubbles blown at $\Qair = 4~\milli\liter\per\minute$. 

The bubble lifetime $\tau$ is plotted as a function of TTAB concentration $c$ in Figure \ref{fig:moyennes_rupture}b, where data points are averages over 6 to 20 bubbles and errors bars show the associated standard deviation. The average bubble lifetime increases with surfactant concentration up to the cmc and seems to saturate around $\tau \sim 80~\second$. The longer lifetime of concentrated bubbles is consistent with our observation that the tangential stress increases with surfactant concentration, hence a slowed down drainage (see Paragraph \ref{sec:comp_exp_drainage}). 

At low surfactant concentrations ($c<\cmc$), the bubble lifetime is relatively scattered, in agreement with observations reported in the literature on very dilute solutions \cite{Lhuissier2011, Modini2013}. Bubble lifetime reproducibility seems to improve above the cmc. Again, we do not observe any significant discrepency between the data sets corresponding to the experiments performed in different conditions (in a glove box, with the needle left in place or removed after generation).
%
\subsection{Discussion}
%
The experimental results presented in Paragraph \ref{sec:bursting_position} have highlighted two distinct behaviours, depending on the surfactant concentration $c$:
\begin{itemize}
\renewcommand\labelitemi{$\bullet$}
\item for $c>\cmc$, the rupture nucleation point is reproducible and located at the bubble apex ;
\item for $c<\cmc$, the bubble nucleation point is rather scattered, whatever the precautions taken, but is getting closer and closer to the bubble foot when the concentration is decreased.
\end{itemize}
The bursting of concentrated bubbles at the apex is quite intuitive, since the film is expected to rupture where it is the thinnest, which is \textit{a priori} at the bubble top because of liquid drainage towards the bubble foot. The transition towards the low concentration regime ($c<\cmc$, with $\Theta$ coming closer to $0$) is in agreement with the results of Lhuissier \& Villermaux \cite{Lhuissier2011} and Modini \textit{et al.} \cite{Modini2013} who respectively observed that tap water bubbles and salty bubbles with traces of surfactant ($<0.1~\cmc$) preferentially burst close to their foot.

However, Debr\'{e}geas \textit{et al.} \cite{Debregeas1998} noticed that viscous silicone oil bubbles -- \textit{i.e.} in the absence of surfactant -- burst at the top. Thus, there is a discontinuity between the behaviour of bubbles made of pure liquid (silicone oil for instance) and bubbles containing minute quantities of surface active agents (like tap water), even though the flow profile is similar in both situations. We indeed showed in Paragraph \ref{sec:comp_exp_drainage} that the flow profile in the bubble cap is getting closer and closer to a plug flow (which is the limit expected with stress-free interface) when the surfactant concentration is decreased. This evidences that the position of the bursting point is no longer controlled by gravity-driven drainage at low surfactant concentrations. Note that the high viscosity of silicone oil bubbles may help damping thickness fluctuations and thus favour drainage-controlled rupture.

\begin{figure}
\centering
\includegraphics[width=6cm]{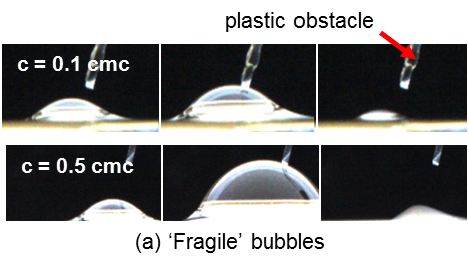}
\includegraphics[width=6cm]{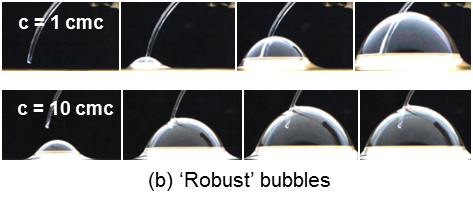}
\caption{The sensitivity of bubbles to an external perturbation is tested by placing a small plastic obstacle next a bubble in continuous generation. Depending on the TTAB concentration, the bubble either (a) bursts when it touches the obstacle or (b) goes through it without damage.}
\label{fig:sequences_fragilite}
\end{figure}

Lhuissier \& Villermaux have attributed the rupture of tap water bubbles to a marginal regeneration, which produces thinner film patches in the vicinity of the bubble foot. However, careful observation of our TTAB-stabilized bubbles reveals that marginal regeneration cells are always present, regardless of the surfactant concentration (see the color image sequence in Figure \ref{fig:generation}b, corresponding to $c=10~\cmc$). This suggests that both rupture factors -- gravitational drainage and marginal regeneration -- coexist for all surfactant concentrations, the former being dominant at high surfactant concentrations and the latter governing the bursting at low concentrations. The dispersion of data points observed for $c<\cmc$ may then be explained by the transition between those rupture mechanisms. \bigskip \\
\indent The ``fragility'' of bubbles may also contribute to the data dispersion at low surfactant concentrations. Indeed, we already noted that drainage measurements at $c<0.8~\cmc$ were impeded by the presence of the plastic cylinder, which caused the bubbles to burst prematurely. In order to demonstrate this effect in a more systematic way, we place a small plastic obstacle (dipped into the surfactant solution beforehand) in the way of inflating bubbles of various concentrations. The image sequences displayed in Figure \ref{fig:sequences_fragilite} show that bubbles made from dilute solutions ($c=0.1$ et $0.5~\cmc$, Figure \ref{fig:sequences_fragilite}a) burst upon contacting the obstacle, while bubbles made from more concentrated solutions ($c=1$ et $10~\cmc$, Figure \ref{fig:sequences_fragilite}b) penetrate it without damage.

For $c<\cmc$, the bubbles are thus ``fragile'', in the sense that they are more sensitive to external perturbations (the obstacle in Figure \ref{fig:sequences_fragilite} but also dust particles for example) than their more concentrated counterparts. This fragility may be attributed to a lower value of surface dilatational elasticity, leading to less efficient healing of surfactant coverage inhomogeneities, and probably contributes to the dispersion of data points at low surfactant concentrations.
%
%
\section{Conclusion} \label{sec:conclusion}
%
In this paper, we have studied how the drainage and bursting of bubbles floating at the surface of a soapy bath depend on the concentration of the cationic surfactant TTAB in water. We have shown that the decrease of the bubble cap thickness at the apex can neither be rationalised by a drainage model with rigid liquid/air interfaces, nor by assuming that the interfaces are stress-free. We proposed a simple model describing the gravitational drainage of a bubble with an intermediate boundary condition at the interfaces (\textit{i.e.} partially rigid interfaces), where the difference with respect to the rigid case is quantified by the extrapolation length $b$. It turned out that the plug flow contribution to the drainage dynamics decreases when the surfactant concentration is increased. This change is likely related to the rise of a surface stress due to the increased presence of surfactants at the interface. A more advanced model, in the spirit of the one developed by Hermans \textit{et al.} \cite{Hermans2015} for the drainage of hemispherical films coated onto a solid lens, could include explicitely Marangoni and surface viscous stresses, which are implicitely encompassed in the extrapolation length $b$ in our approach.

The position at which a hole nucleates in the bubble cap was found to depend strongly on the surfactant concentration. Bubbles made from concentrated solutions ($c>\cmc$) systematically burst from the apex, while the rupture nucleation point was observed to be quite scattered at low concentrations ($c<\cmc$), but getting closer and closer to the bubble foot on average. This low-concentration behaviour is correlated with an increasing fragility of the bubbles with respect to perturbations. Building on the ideas of Lhuissier \& Villermaux \cite{Lhuissier2011}, we suggested that both gravitational drainage and marginal regeneration are involved in the bursting of bubbles, the former prevailing at high surfactant concentration and leading to a hole nucleation at the apex, and the latter being responsible for bursting at the bubble foot at low concentrations. Besides, we observed that marginal regeneration is present at all of the concentrations that we probed, but it seems to become a weaker contribution to bursting as the concentration is increased. The fragility of low-concentrated bubbles with respect to perturbations may be a clue to explain why marginal regeneration driven bursting occurs only at low surfactant concentrations. These results also open the question of the relation between the properties of the marginal regeneration patches and the physicochemical properties of both the surfactant solutions and the surfactant molecules themselves.
%
\section*{Acknowledgments}
%
The authors would like to thank Emeline Alvarez for her help with the bubble bursting experiments. L.C. was supported by ANR F2F and this work also benefited from the support of an ERC Starting Grant (Agreement 307280-POMCAPS).
%
\bibliography{biblio}
\bibliographystyle{unsrt}
%
\end{document}